\documentclass[10pt,preprint]{aastex}
\usepackage{graphicx}

\begin{document}

\title{\bf A concentration of quasars around the jet galaxy NGC1097}

\author{H. Arp}
\affil{Max-Planck-Institut f\"ur Astrophysik, Karl Schwarzschild-Str.1,
  Postfach 1317, D-85741 Garching, Germany}
 \email{arp@mpa-garching.mpg.de}

\author{D. Carosati}
\affil{Armenzano Astronomical Observatory, 06081 Assisi(PG), Italy}

\begin{abstract}

A quasar search in the region of the active galaxy NGC 1097
yielded 31 quasars in 1984. After completion of the 2dF survey in 2004
the number of catalogued quasars just within 1 degree of the galaxy
increased to 142. About 38 $\pm$ 10 of these are in excess of average
background values.

The evidence in 1984 is confirmed here by an increasing density of quasars
as one approaches NGC 1097. Quasars within 1 degree differ from the
background by being significantly brighter. There also appear two
elliptical rings or arcs of quasars at r $\sim$ 20'and 40'.    

\end{abstract}


\section{INTRODUCTION}
 
The galaxy NGC 1097 has long been known as an unusually bright Seyfert
with long, straight jets emerging from its active nucleus (Wolstencroft
and Zealey 1975; Arp 1976; Lorre 1978). Analysis of X-ray emission
shows connections with quasars (Arp 1999). The nucleus of NGC 1097 
has been much studied as the site of strong X-ray, radio and optical
emission (see for example Nemmen et al. 2006). Wolstencroft, Perley
and Tully (1984) interpreted the jets in terms of thermal
bremstrahlung. But varying opinions have been expressed as to the
nature of these exactly straight, low surface brightness features 
(Higdon and Wallin 2003; Wehrle et al. 1997). Fig. 1 here, shows the
stacked, low surface brightness enhanced, image of NGC 1097 from CTIO
4 meter plates taken by Arp (Lorre 1978).

Adding to the interest in this exceptionally active Seyfert
galaxy, a high density of a small number of bright quasars near NGC
1097 was found by Wolstencroft et al. (1983). Shortly thereafter two
objective prism plates were taken at the UK 1.2 meter schmidt
telescope. Mr. X. T. He identified lengthened spectra of 104
quasar candidates in the central 8.1 sq.deg. of NGC 1097. The most
likely of these were spectroscopically confirmed by redshifts observed
by Arp at the 2.5 meter Du Pont telescope in Las Campanas, Chile.
 
Well determined redshifts of 31 quasars to slightly
fainter than V = 20.0 magnitude then became available and their
distribution with respect to NGC 1097 was analyzed in Arp et
al. (1984). There Figs. 5, 9, and 10 demonstrated the excess of quasars
around NGC 1097 compared to the background. Figs. 11 and 12 showed the
distribution of these quasars increased in density with decreasing
distance to the galaxy. From a $\chi^2$ test the chance of this
observed distribution being consistent with a uniform distribution
turned out to be extremely small (0.008). Fig. 12 from that 1984 paper
shows most clearly the two peaks in quasar distribution for the
quasars with V $\leq$ 19.5 which is confirmed here. 

\newpage

\begin{figure}[ht]
\includegraphics[width=7.0cm]{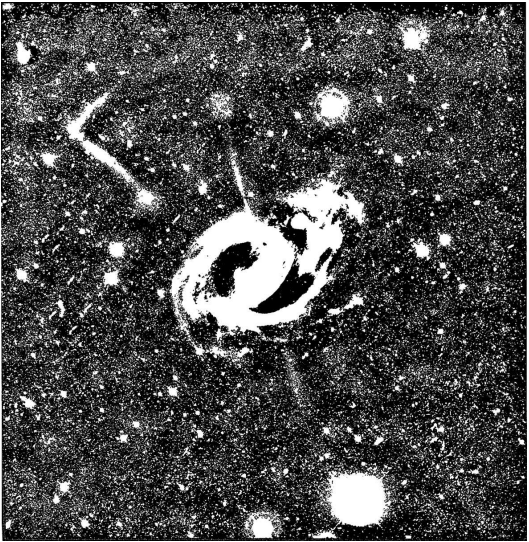}
\caption{Faint surface brightness jets in NGC 1097 (30x30')
\label{fig1}}
\end{figure}

\begin{figure}[ht]
\includegraphics[width=10.0cm]{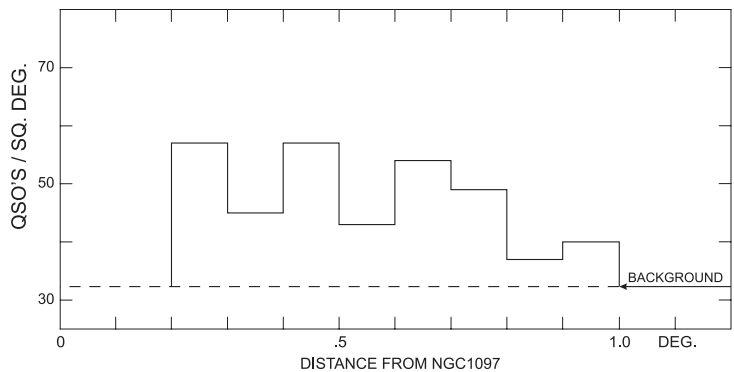}
\caption{Density of 142 quasars in concentric annular rings around
NGC 1097.
\label{fig2}}
\end{figure}

\section{New Quasars from the 2dF Survey}

When the 2dF survey was completed (Croom et al. 2004; Smith et
al. 2005) a homogeneous survey of quasars down to apparent magnitude
$b_J$ = 20.85 mag. became available for two strips across the sky of 5 
degrees width in declination. Serendipitously, NGC 1097 at Dec. =
-30deg. 17 min., was closely centered in the southern strip.
The important question then arose as to how many of this large number
of quasars around NGC 1097 were in excess of expected background counts.
We calculate here the excess and its significance in two ways.

1) The 2dF survey quotes 23,338 QSO's in a total survey area of 721.6
   sq. deg. for an average density of 32.34 QSO's/sq.deg. Within a
   circle of radius 1 deg. that would predict 101.6 QSO's as an average
   background.

   The brightest jet reaches out past the nearest quasar at r
   $\sim$ 11'. Inside of this there is evidence for considerable
   absorption and if we take the area from which NGC 1097 blocks out 
   background quasars it is .10 sq. deg. A minimum of about 3.4
   quasars would be obscured or not identified. We could then
   calculate the excess of quasars over normal background as: 

   (142 + 3.4 - 101.6) = 43.8/$\sqrt101.6$ =  4.3 sigma

2) However it might be objected that six of the 142 NED quasars within   
   1 deg. of NGC 1097 are not 2QZ objects. They are bright apparent
   magnitude (17.1 to 19.5 mag.) mostly discovered in the
   homogeneous, wide field search in 1984. Perhaps from extended
   blockage by the galaxy or crowding by bright stars the 2dF has
   missed them. Nevertheless, leaving out these six still gives a
   significant excess of NGC 1097 quasars of:  

   (136 + 3.4 - 101.6) = 37.8/$\sqrt101.6$ = 3.8 sigma.

   We adopt the second result as the more conservative but note that
   the reddening/absorption extending over the furthermost jets 
   probably obscures more than we have assumed and therefore raises
   the counts appreciably (Arp 1999).

   In the southern 2dF strip the region most clear of blank squares is
   west of NGC 1097. The average of seven 1 deg. radius samples of only
   2QZ quasars give 108 $\pm 8.5$ (p.e). compared to the 101.6 value
   for the entire 2DF. The quasar distribution is not strongly non
   random.The large signals then give probabilities which are
   significant. As the sample size increases so does the tolerance for
   deviations from the Gaussian.

\section{Increase of quasar density near NGC 1097}
  
Taking the 142 quasars discovered in uniform surveys covering the area
within a radius of 60' we can ask the very important question: ``How
are these quasars distributed in radial distance from NGC 1097?''
Figure 2 here shows that the density of quasars is conspicuously above
background from about 12 arcmin to about 48 arc min from the galaxy
and then begins to fall until at about 1 degree it joins the expected
background counts. Such a consistent increase toward the galaxy of a
large number of objects argues that the association with NGC 1097 is
not accidental.  

Secondly there are apparent rings at about 20 and 40 arcmin. The same
peaks in the radial distribution showed clearly in the brighter
quasars of Figs. 11 and 12 of Arp et al.(1984). Circles and arcs have
been observed around other active galaxies (UGC 8584 and others as
reported in Fulton and Arp 2007, in preparation). They could be a
consequence of multiple ejections of shells of matter. 

The association is further supported by quasars which show an
azimuthal concentration along the lines of the NE and SW optical jets.
These concentrations are shown particularly well in the bright
quasars in Fig. 13 of the 1984 paper on NGC 1097. One can also see
this in Fig. 3 here, where the string of 9 quasars is seen crossing
the outer ellipse in the NE direction.

\begin{figure}[ht]
\includegraphics[width=14.0cm]{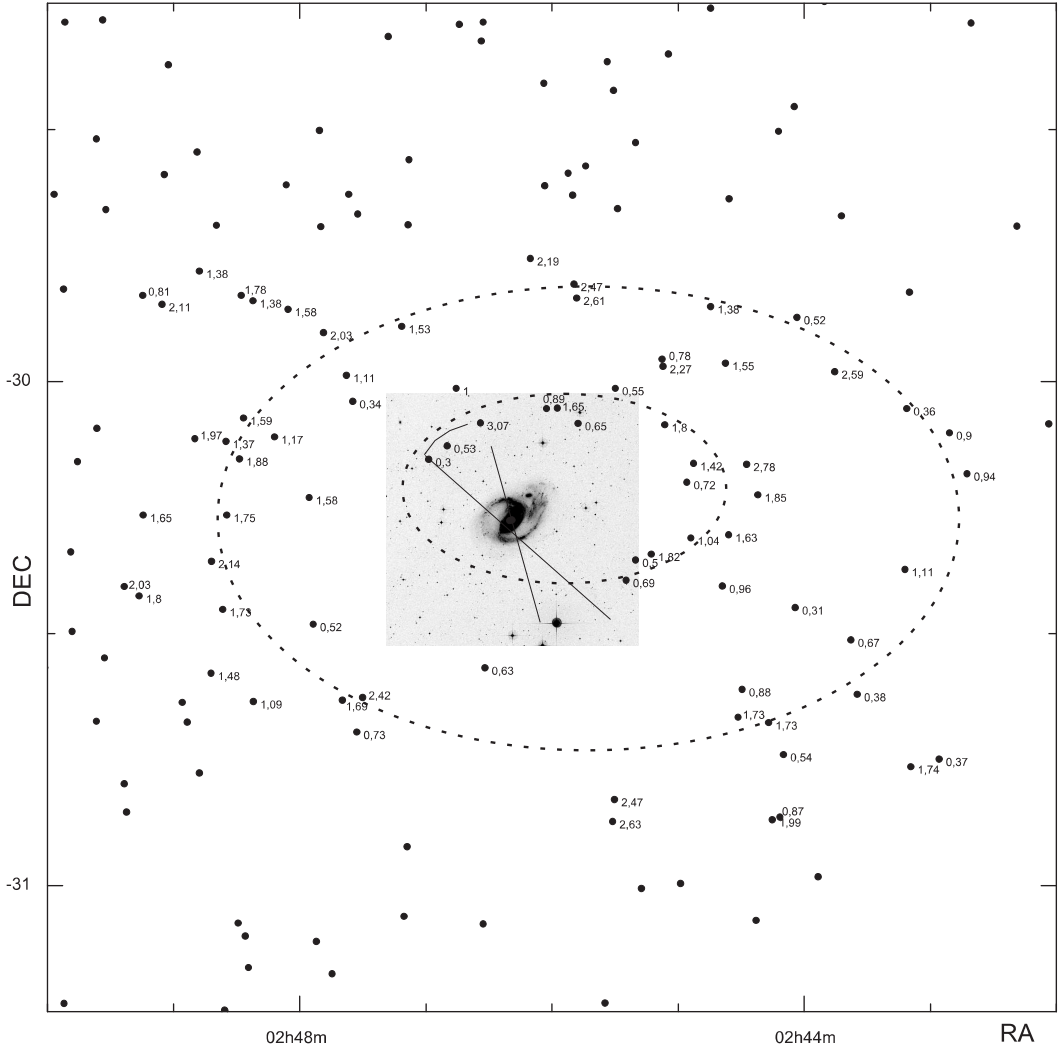}
\caption{Distribution of catalogued quasars around NGC 1097.
\label{fig3}}
\end{figure}

\begin{figure}[ht]
\includegraphics[width=15.0cm]{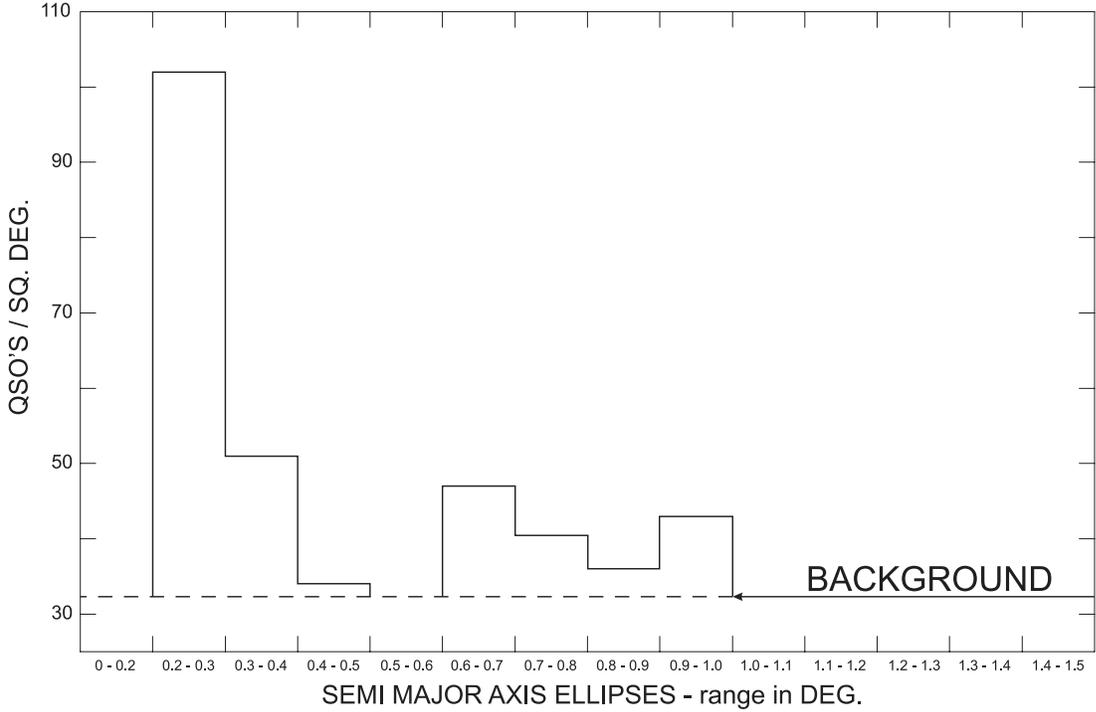}
\caption{Density of quasars in concentric, elliptical annular rings
around NGC 1097.
\label{fig4}}
\end{figure}

\newpage

\section{Distribution of Quasars in rings and ellipses}

The plot of quasars in Fig. 3 shows that there is an inner
distribution of about 18 quasars which fill out an ellipse which is
cleanly defined except in the southern portion where the jets become
very faint and red - just in the direction in which Arp (1999)
concludes there is overlying absorption. If rings and arcs are the
result of ejection in a plane it would in general be expected to view
the features at an angle. So ellipses could be common. Observationally
the ``rings'' seen in the radial distributions of Fig. 2 are quite
broad but, fitted to the Fig.3 ellipses they are remarkably sharp as
shown in Fig4.

They perhaps suggest an event or epoch of ejection. Note the curious
double quasars N and S in the outer ring/ellipse with redshifts z =
2.47 and 2.61 opposite a pair with 2.47 and 2.63. Also a pair NW - SE
at z = .52 and .52. 

\newpage

\begin{figure}[p]
\includegraphics[width=16.0cm]{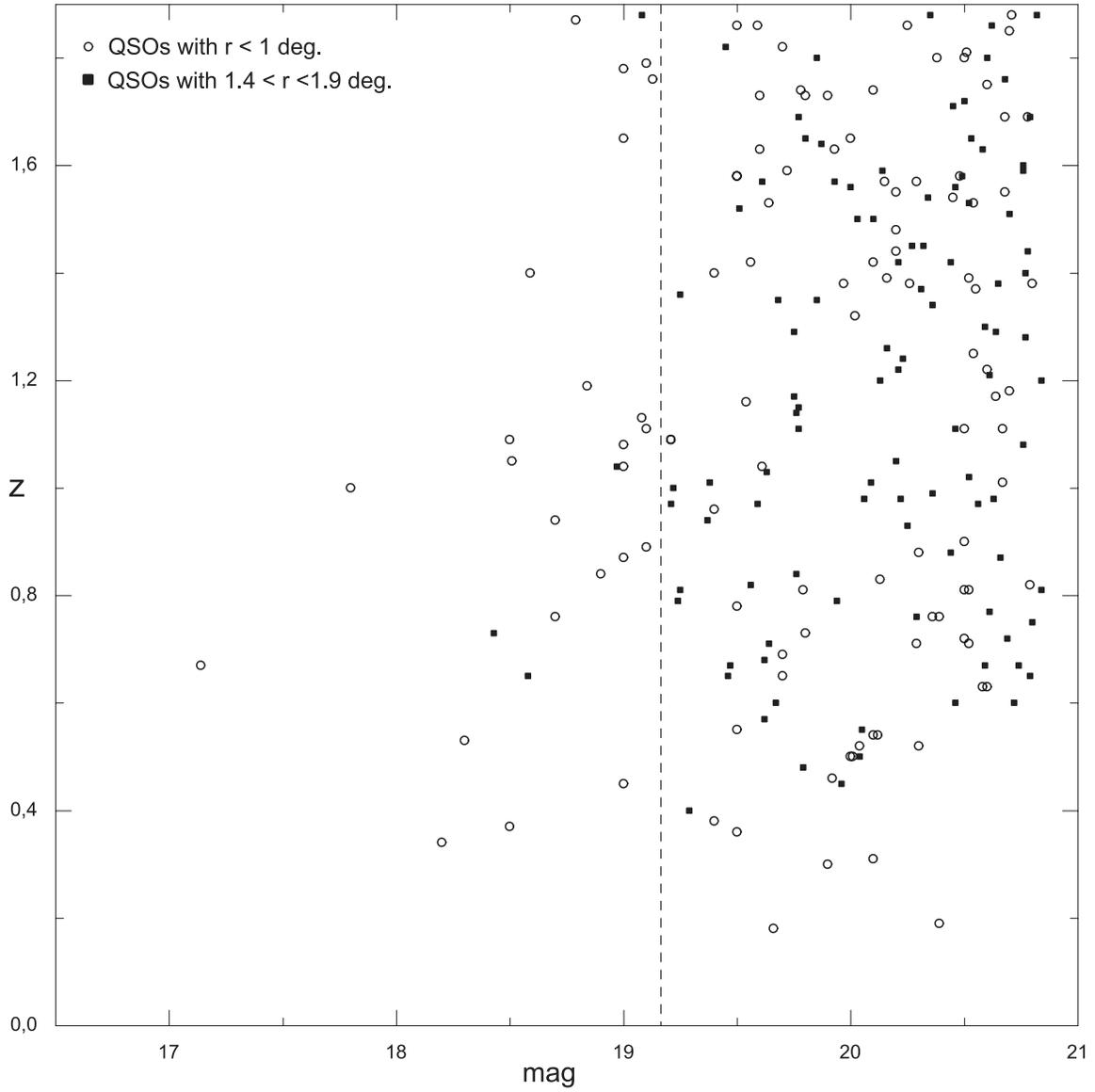}
\caption{Redshift, z, vs apparent magnitude. Open circles are QSOs
$\leq$ 1 deg from NGC 1097. Filled squares are from outer annulus 
between 1.4 and 1.9 deg. Limit to bright sample at 19.16 mag.
\label{fig5}}
\end{figure}

\section{Another way to test associations}

Fig. 5 shows that the apparent magnitudes of the quasars within 1
deg. of NGC 1097 are systematically brighter than the quasars in a
control ring between 1.4 and 1.9 deg. For QSOs with z $\leq$ 1.9 and
mag. $\leq $19.6 mag., there are 24 which are found near NGC 1097
compared to 4 found in the more distant comparison ring. There were
142 QSOs around NGC 1097 and 147 in the ring. The area in the annular
ring is 5.18 sq. deg. and the area within 1 deg. of NGC 109 is 3.14
sq. deg. Therefore the observed number of bright, field QSOs should be
reduced by $4 x 142/147 x 3.14/5.18 = 2.34$ on just area considerations. 

But restricting the test to only the sample pictured in Fig. 5 we can
say:  If the 24 + 4 bright quasars were an even mixture from around NGC 1097
and from the ring control field, then the chance of accidentally
choosing 24 which were around the galaxy would be $^{24}P(.5)_{28} = 7.6 x
10^{-5}$. Even if we left out 4 QSOs in this range which 2dF did not
detect we would still have $^{20}P(.5)_{24} = 2.9 x 10^{-4}$. Correcting  
for the larger area covered in the ring would lessen the latter
probability by about another factor of 7.

\newpage

\section{Conclusions}

In 1984 and again in 1999 evidence was presented that a number of
quasars were physically associated with the very bright, active
Seyfert Galaxy NGC 1097 (Arp et al. 1984; Arp 1999.) Twenty years
later this evidence is confirmed by a homogeneous survey with modern
detection techniques which accidently happened to cover NGC 1097 and
adjoining regions. Of the order of 38 quasars have been found in
excess of background and in high density concentration around the
galaxy. 

It has been clear for all this time that the origin of quasars must
have something to do with the central galaxy. Since it is
evident from early investigations that NGC 1097 is ejecting material
into its surrounding regions, it would be an obvious conclusion
that these quasars have been ejected from the active nucleus of the 
Seyfert galaxy. For example, although the nature of the optical jets
is still uncertain, study of the 4 meter CTIO direct images
showed where ejected material associated with the jets had broken
through the spiral arms of the galaxy (Arp 1976). Some material has
clearly come out of the nucleus. 

In the intervening years there has been much further evidence for
such an origin for quasars (see e.g. Arp 1998; 2003). The greatest number of
quasars aligned across galaxies was six for NGC 3516 (a Seyfert) and
five for NGC 5985 (another Seyfert). Now we have a much larger number,
of the order of 38, associated with one of the brightest and most
active Seyferts known, NGC1097. The question then arises as to whether
all quasars have origins in relatively nearby galaxies.

In a very complete and informative discussion, of their own and 23
other analyses, Nollenberg and Williams (2005) establish that there are
strong quasar cross correlations on the sky with galaxies up to 110
arcmin separation. They note qualitative agreement with gravitational
lensing models but point to the need for much greater quantities of
cold dark matter (CDM). They also mention a few cases of ``physical
associations'' and the possibility of ``something much more exotic
such as photon decay.''
                                                         
NGC 1097 represents an approach through studying quasar 
associations with individual active galaxies. From the many cases so
far found the obvious generalization is to all quasars. As for theory,
for 13 years there has been a solution for elementary particles with
increasing mass (Narlikar and Arp 1993) which yields the general case of 
greater redshift for younger objects. However, even if there were only
a few cases of intrinsic redshifts of quasars one would have to consider
continuity of physical properties between quasars and galaxies as
requiring the current fundamental assumption about redshift distances in
general to be reexamined.

\subsection{Epilogue}

It was commented on the title page of this web posting that the paper had
been rejected by the Astrophysical Journal Letters. Thus the editor
spake: ``Your paper has not been able to convince either of two
independent referees. . . . ``No suppression of your work has been done
through my action since you are welcome to submit your paper to a
different journal.''
 
The information supplied here should enable the readers to decide for
themselves the value of the data and its discussion. But perhaps more
important it enables a judgment on the core structure of current
science. 

\section{References}

\noindent

Arp, H. 1976, ApJ 207, L147

Arp, H. 1998, ``Seeing Red'', Apeiron, Montreal

Arp, H. 1999, Ap\&SS 262, 337

Arp, H. 2003, ``Catalog of Discordant Redshift Associations'', Apeiron,
Montreal 

Arp, H., Wolstencroft, R., He, X.T. 1984, ApJ 285, 44

Croom, S., Smith, R., Boyle, B., Shanks, T., Miller, L.,
Outram, P., Loaring, N., 2004, M.N.R.A.S. 349, 1397

Higdon, James L., Wallin, John F.  2003, ApJ 585, 281

Lorre, J. 1978, ApJ, 222, L99

Narikar, J., Arp, H. 1993 ApJ 405, 51 

Nemmen, R., Storchi-Bergmann, T., Yuan, F., Eracleous, M., Terashima,
Y., Wilson, A. 2006, ApJ 643, 652 

Nollenberg, J., Williams, L. 2005 ApJ 634, 793

Smith, R. Croom, S., Boyle, B., Shanks, T., Miller, L., Loaring,
S. 2005, M.N.R.A.S. 359, 57

Wehrle, A., Keel W., Jones, D. 1997, AJ 114, 115

\end{document}